# Bridging the gap between nanowires and Josephson junctions: a superconducting device based on controlled fluxon transfer across nanowires

E. Toomey[1], M. Onen[1], M. Colangelo[1], B. A. Butters[1], A. N. McCaughan[2], and K. K. Berggren[1,*]

[1]*Massachusetts Institute of Technology, Department of Electrical Engineering and Computer Science, Cambridge, MA 02139, USA*

[2]*National Institute of Standards and Technology, 325 Broadway, Boulder, Colorado 80305, USA*

*correspondence to: berggren@mit.edu

**ABSTRACT**

The basis for superconducting electronics can broadly be divided between two technologies: the Josephson junction and the superconducting nanowire. While the Josephson junction (JJ) remains the dominant technology due to its high speed and low power dissipation, recently proposed nanowire devices offer improvements such as gain, high fanout, and compatibility with CMOS circuits. Despite these benefits, nanowire-based electronics have largely been limited to binary operations, with devices switching between the superconducting state and a high-impedance resistive state dominated by uncontrolled hotspot dynamics. Unlike the JJ, they cannot increment an output through successive switching, and their operation speeds are limited by their slow thermal reset times. Thus, there is a need for an intermediate device with the interfacing capabilities of a nanowire but a faster, moderated response allowing for modulation of the output. Here, we present a nanowire device based on controlled fluxon transport. We show that the device is capable of responding proportionally to the strength of its input, unlike other nanowire technologies. The device can be operated to produce a multilevel output with distinguishable states, which can be tuned by circuit parameters. Agreement between experimental results and electrothermal circuit simulations demonstrates that the device is classical and may be readily engineered for applications including use as a multilevel memory.

**1. INTRODUCTION**

Superconducting devices have played critical roles in technologies such as quantum computing[1], astronomical imaging[2,3,4], magnetometry[5,6], and digital logic[7,8,9,10]. Past superconducting electronics have largely been based on the Josephson junction (JJ) due to its desirable characteristics, such as rapid operation speeds >100 GHz and power dissipation on the order of $10^{-19}$ J per switch[11]. Recently, however, superconducting nanowires have emerged as an alternative platform for new electronics. Unlike JJs, superconducting nanowires are dominated by thermal dissipation and a loss of phase coherence, switching from a superconducting state to a high-impedance resistive state when triggered by an external signal, such as a photon or current pulse. This functionality has enabled nanowires to be used in applications where JJs fall short—for instance, the nanocryotron (nTron) is a comparator-like nanowire device that can support high impedances and large fanout[12], whereas JJs lack intrinsic device gain and compatibility with high-impedance environments. The nTron can also be triggered by SFQ pulses to interface between JJs and CMOS circuits[13,14], demonstrating that nanowire electronics have a unique place in computing architectures. Other advantages of nanowires including their single layer fabrication, high output voltages, and scalability have prompted the development of new devices for use in readout, memory, and sensing, while novel two-dimensional superconducting structures have also recently been explored as a basis for studying correlated physical phenomena[15].

Despite the growth of nanowire electronics, further advancement into new applications is currently hindered by characteristics that are inherent to nanowires. By switching from the superconducting domain to the resistive domain in response to an external input, nanowires are limited to operations of two states, with the resistive state expelling nearly all current from the high-impedance hotspot and generating a single large output. For the nTron, this means that the output voltage is fixed and only dependent on the gate voltage exceeding a critical threshold. Similarly, a recently reported superconducting nanowire memory based on persistent current was limited to binary operations of either "0", no current stored in the superconducting loop, or "1", a maximum amount of current stored in the loop[16]. By breaking coherence, hotspot formation also prevents incrementation of the output by successive device switching, as can be done with JJs. Another disadvantage of nanowire devices is that the operation speeds are slow due to the long thermal healing time of the resistive domain, which is often made slower by the electrical time constants of the biasing circuit.

These limitations leave a gap in the family of superconducting electronics for a device that offers the robust interfacing capabilities of a traditional nanowire, but with a more moderated response that allows for modulation of the output, analogous (but not identical) to incrementing in a JJ. Past approaches to creating such a device have focused on Dayem bridge weak-links[17,18], where the dimensions of a nanowire in relation to the material's coherence length allow for preservation of phase coherence over a temporary phase-slip center. However, there have been very few experimental demonstrations of these devices in real circuit operations, and the primary goal has largely been to achieve true Josephson behavior in a nanowire rather than to demonstrate a device with intermediate characteristics of both technologies.

Here, we report on a superconducting nanowire device based on thermal principles that demonstrates a controlled output. Unlike other nanowire technologies, the device responds proportionally to the magnitude of an input signal and can be operated to achieve multiple discrete states. By using local resistive shunting and a high-inductance superconducting loop, we are able to controllably trap flux[1] in quantities of $n\Phi_0$, where $n$ is an integer less than 10, and $\Phi_0$ is the magnetic flux quantum. We experimentally show that the amount of flux per event $n$ is dictated by circuit parameters, and validate these results with electrothermal simulations. The results show that the device may be designed to achieve different $n\Phi_0$ outputs depending on the desired application. We anticipate that this device will serve as a foundation for new nanowire technologies such as a multilevel memory or multilevel logic circuit elements.

---

[1] In this work, we will interchangeably use the terms "flux" and "fluxoid" for the sake of simplicity; however, it should be clear that we always mean "fluxoid", as trapped flux in a superconducting loop manifests itself in quantized circulating current, or fluxoids.

## 2. DEVICE CHARACTERIZATION

Figure 1 summarizes the device architecture and its basic characteristics. The device is comprised of three superconducting nanowire elements: a narrow constriction, a storage loop, and a readout tool known as the yTron[19]. All three elements were fabricated together on a ~20 nm thick niobium nitride (NbN) film on a silicon oxide substrate using electron beam lithography. In addition to the nanowire components, a resistive metal shunt was patterned in parallel with the constriction to reduce Joule heating and provide damping, similar to the purpose served in resistively shunted JJs. Initial efforts to fabricate the resistor on top of the NbN film failed due to contact resistance, requiring us to place the resistor layer beneath the NbN film, as was done in similar work on Nb nanowires[20]. Previous attempts to shunt nanowires have found that series inductance between the shunt and the constriction plays a critical role in the effectiveness of the damping; high series inductance produces relaxation oscillations, while increasing the inductance even further makes the resistor completely ineffective[21]. To reduce this effect, the shunt was patterned as close to the constriction as possible and the leads were made wide to reduce the number of squares of material. Details of the complete fabrication process may be found in the Methods section.

A simple circuit model for the device is shown in Fig. 1b. To trap flux in the loop, a bias current $I_{bias}$ is inductively split to the nanowire in the amount of $\alpha I_{bias}$, where $\alpha = L_{loop}/(L_{constriction} + L_{loop})$. In this case, $L_{constriction} = 284$ pH and $L_{loop} = 1.87$ nH, causing α = 0.87. Once the sum of $\alpha I_{write}$ and any existing current circulating in the loop surpasses the critical current of the constriction, the nanowire switches and the bias current is diverted away from the constriction to the shunt resistor and the righthand side of the loop. By shunting the majority of the bias current, the resistor allows the nanowire to recover the superconducting state more quickly and limits the amount of current that charges $L_{loop}$, thus controlling the amount of flux that is trapped once the constriction heals. After the constriction heals, a persistent current circulates in the superconducting loop in quantized units of $n\Phi_0/L$, where *n* is an integer, $\Phi_0$ is the magnetic flux quantum, and *L* is the total loop inductance. Since the geometric inductance of this device is < 0.33 fH, the total loop inductance is dominated by the kinetic inductance of the superconducting nanowires. In this device, persistent current is estimated to be quantized in approximately 0.95 µA/fluxon.

The amount of circulating current in the loop can be nondestructively read out using the yTron. As described by McCaughan et al.[19], the yTron is a three-terminal device with two adjoining arms whose switching currents depend on one another as a result of current crowding around the intersection point. In our device, the left arm of the yTron forms part of the superconducting loop so that the switching current of the right arm is a function of the amount of circulating current—a higher circulating current in the clockwise direction flowing through the yTron's left arm will result in a higher switching current of the yTron's right arm. Since the two arms of the yTron are electrically disconnected from one another, switching the right arm does not break superconductivity in the left. As a result, the state of the loop is undisturbed by the reading process, allowing us to nondestructively sense the amount of circulating current.

Figure 1c shows the current-voltage characteristics of an isolated shunted nanowire patterned alongside the device with dimensions identical to those of the constriction. The absence of hysteresis, as shown by the lack of separation between the switching and retrapping currents, indicates that the shunt resistor was able to reduce Joule heating through the constriction by effectively diverting the bias current, thereby reducing power dissipation in the nanowire and allowing it to regain the superconducting state more quickly[22]. Observation of the amplified RF output of the device within a bandwidth of 2 GHz did not reveal any relaxation oscillations, suggesting that the shunt inductance was low enough to prevent stable relaxation oscillations at least within the limits of our measurement capabilities[21].

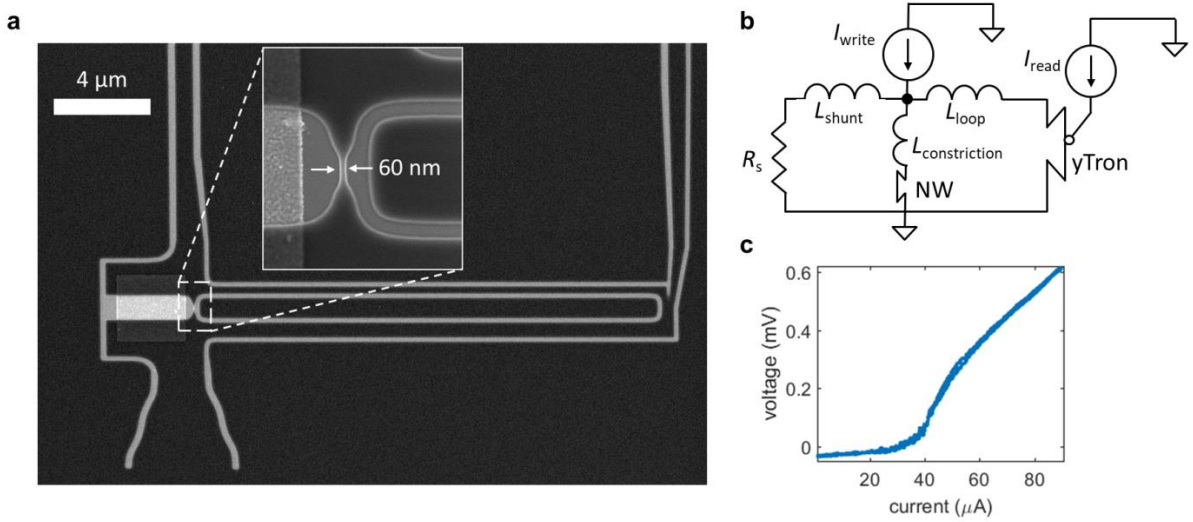

**Figure 1: Device design and characterization.** (a) Scanning electron micrograph of the device. The dark area is the NbN film, while the light outlines are the underlying substrate. The inset shows an enlarged view of the 60-nm-wide constriction in parallel with the resistive shunt. The shunt dimensions were 1 μm x 3 μm. The right-hand side of the loop is connected to a yTron with arm widths equal to 300 nm. (b) Circuit representation of the device. In this particular design, $R_s$ = 5 Ω, $L_{shunt}$ = 50 pH, $L_{constriction}$ = 284 pH, and $L_{loop}$ = 1.87 nH. (c) Current-voltage characteristics of an isolated shunted nanowire of width = 60 nm, $R_s$ = 5 Ω. The absence of hysteresis implies that Joule heating through the nanowire has been significantly reduced by the presence of a shunt resistor. Comparison to the characteristics of an unshunted nanowire may be found in the Supplemental Material.

## 3. DEMONSTRATION OF CONTROLLED DYNAMICS

Figure 2 shows the response of the device to an input voltage pulse of varying amplitude and width; the response is compared to that of an otherwise identical device lacking a resistive shunt. For these measurements, an input voltage of width ranging from 5 ns to 100 μs and height ranging from 50 mV-550 mV was sent to the constriction through a -30 dB attenuator. The amount of stored current in the loop was inferred by measuring the switching current of the yTron. Before each positive input pulse, a negative pulse (width = 10 μs, height= -1.3 V) was sent to the constriction to reset the superconducting loop. Details of the experimental setup may be found in the Methods section.

As shown in Fig. 2a, the amount of stored current in the unshunted device sharply increased with increasing input voltage, but then abruptly dropped off, suggesting instability. This response was also observed in the device reported in Ref.[23], and was speculated to be due to overheating of the constriction, causing flux to be lost. In contrast, the response of the shunted device in Fig. 2b shows that the amount of flux stored in the loop increases proportionally with input voltage. Unlike the unshunted constriction, in the shunted device there was no sudden loss of stored flux or signatures of unstable oscillations, implying that heating in the constriction was moderated by the presence of the resistive shunt.

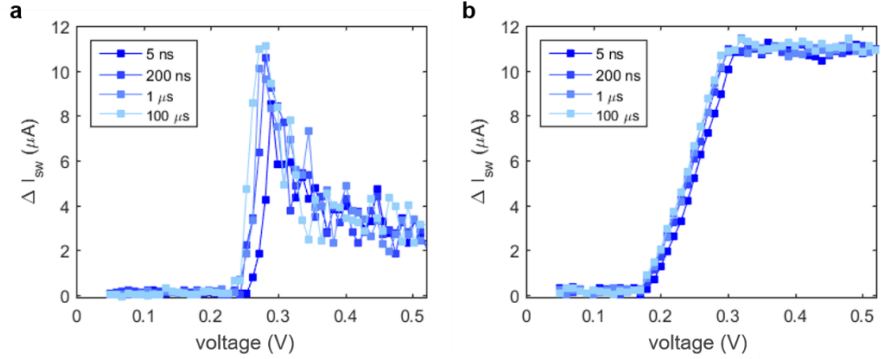

**Figure 2: Response to input signal voltage and pulse width.** (a) Switching current of the right yTron arm as a function of the input voltage for a superconducting loop with an unshunted constriction. The input pulse widths ranged from 5 ns to 100 μs. The switching currents are plotted in terms of $\Delta I_{sw} = I_{sw} - I_{sw}(v = 0)$, where $v$ is the voltage height of the input pulse. The decrease in $\Delta I_{sw}$ suggests that some flux has been lost in the loop at high input voltages, potentially due to overheating of the constriction. Each point represents the mean of 10 sequential measurements of the yTron switching current. (b) Results from the same measurement repeated on a superconducting loop with a shunted constriction, $R_s = 5$ Ω. The proportional increase in switching current with increasing input voltage implies that the dynamics of the constriction are controlled by the presence of the shunt resistor. Logarithmic colormaps showing the complete range of input pulse widths and voltages for both devices may be found in the Supplemental Material (see Fig. S7).

To demonstrate the flux shuttling capabilities of the shunted device, we measured its dependence on the previously written state by ramping a DC bias current on the constriction without resetting the loop, and recording the switching current of the yTron at every bias point. Details of the measurement setup may be found in the Methods section. As shown in Figure 3a, ramping the bias current to the constriction produced either increasing or decreasing steps in the switching current output of the yTron, signifying a sudden addition or subtraction in the amount of trapped flux. The horizontal lines in Fig. 3a show that the steps could be categorized into seven distinguishable states, revealing that successive switching of the constriction produced controlled, incremental changes in the amount of circulating current in the loop, rather than storing the maximum amount of current every time. The slight variation in the position of the seven states occurred due to instability in the plateaus, representing when the loop current was nearly maximized ($\sim |I_{c,NW}|$) and may have lost a small amount of flux (1-2 $\Phi_0$) to achieve stability. Despite the small shifts at the plateaus, the seven states had well-separated mean values including consideration of their standard deviations (see Supplemental Figure S8). In contrast, Fig. 3b displays the results from repeating the measurement on an unshunted device of the same geometry. In this case, no intermediate states were observed, and the loop trapped nearly its maximum amount of circulating current whenever the nanowire switched. Thus, it was not possible to achieve distinguishable intermediate states without the presence of a resistive shunt.

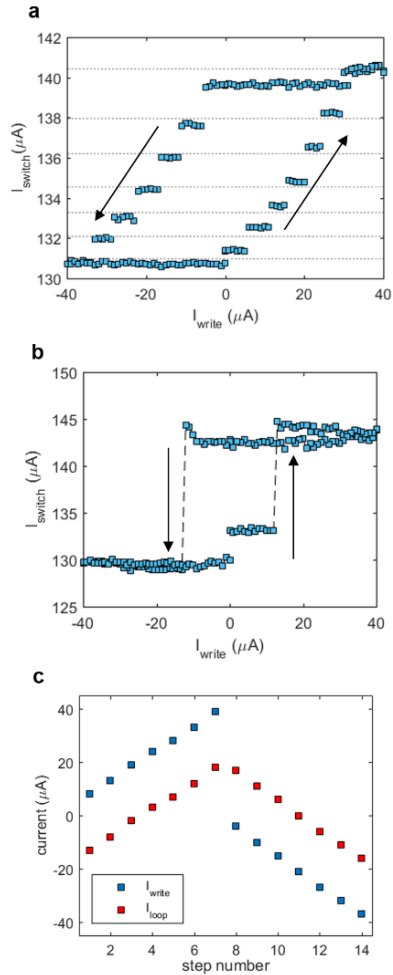

**Figure 3: Demonstration of controlled flux shuttling.** (a) Switching current of the right arm of the yTron on the shunted device in response to a DC bias ramp of ± 40 μA. Each point is the mean of 10 measurements of the yTron switching current. Each bias current step was applied to the constriction for 100 ms before being turned off during the reading operation. (b) Repetition of the same measurement on a device with an unshunted constriction. In comparison to the shunted device, no repeatable intermediate states are observed, and nearly the maximum amount of current is trapped every time the constriction switches. (c) Blue squares represent the bias current used to write to the constriction at the initial point of each of the first 14 steps of the yTron switching current in (a). The approximate circulating loop current (red squares) is also calculated using $I_c = 21$ μA. The data shows that the amount of circulating loop current can be incremented in nearly even steps of ~ 5 μA.

While the yTron is an effective tool for sensing when there is a change in circulating current, it can be imprecise for extracting the exact amount of circulating current in the loop. Depending on the geometry of the yTron, there may be a nonlinear relationship between the amount of circulating current and the induced change in switching current (see Supplementary Figure S6). Additionally, the sensitivity of the yTron depends on the intersection point between its two yTron arms, which has a radius of curvature < 5 nm, leaving room for fabrication variability and thus differences in sensitivity between yTrons of identical design.

To bypass this shortcoming, we used the yTron only to sense *when* a change in trapped flux occurred, and examined the corresponding bias current at each of the points of change in order to infer the magnitude of the loop current. Figure 3c shows the bias current at each of the first 14 steps of the plot in Fig. 3a. The bias current at each step can be used to estimate the amount of circulating current remaining in the loop, given that the transition occurs when the nanowire switches, or when $|\alpha I_{write} + I_{loop}| > |I_{c,NW}|$. The average zero flux state ($I_{loop} \approx 0$) occurred at $I_{c,NW} \approx 20.48 \pm 1.45$ µA over a set of eight bias ramps. While the seven levels in the yTron switching current of Fig. 3a were spaced unevenly, the steps in terms of bias current occur at nearly equal intervals of ~5 µA, or roughly 5 $\Phi_0$ of circulating current. As a result, it is possible to infer that the loop gains or loses ~ 5 $\Phi_0$ of trapped flux every time the constriction switches. Repeating this measurement over eight ramping cycles with a finer sweep produced an average of 4.77 $\Phi_0$ per step in circulating current, with variation from an integer amount ($n \approx 5$) expected to be caused by noise in the measurement setup. A discussion of the experimental noise may be found in the Supplemental Material. Thus, despite the nonlinear response of the yTron, movement of a controlled $n\Phi_0$ of flux per step could be validated by examining the bias current at which each of the steps occurred.

Table 1 summarizes the result of repeating this measurement on two other devices with varying circuit parameters. Device 2 had $R_s = 7.8$ Ω and $L_{loop} = 1.87$ nH, and Device 3 had $R_s = 7.8$ Ω and $L_{loop} = 0.66$ nH. All other geometries and parameters were kept the same. For each device, the bias current was ramped in the same way as before, and the switching current of the yTron right arm was used to indicate when the amount of circulating current had changed. The results from Table 1 show that increasing $R_s$ decreases the number of consistent states and increases the average number of fluxons trapped per switching event, while decreasing the loop inductance may slightly reduce the amount. Plots showing the states of yTron output for Device 2 and 3 may be found in the Supplemental Material.

| Device | $L_{loop}$ | $R_s$ | # states | $\mu$ | $\sigma$ |
|---|---|---|---|---|---|
| Device 1 | 1.87 nH | 5 Ω | 7 | 4.77 $\Phi_0$ | 1.23 $\Phi_0$ |
| Device 2 | 1.87 nH | 7.8 Ω | 5 | 7.63 $\Phi_0$ | 1.5 $\Phi_0$ |
| Device 3 | 0.66 nH | 7.8 Ω | 3 | 5.87 $\Phi_0$ | 1.02 $\Phi_0$ |

**Table 1: Mean change in circulating current per switching event, represented in terms of flux.**

## 4. ELECTROTHERMAL SIMULATIONS

To better understand how the device parameters influence the amount of flux *n* trapped per switching event, we modeled the dynamics of the system using circuit simulations that included the electrothermal dynamics of the resistive hotspot[24] and material-specific physical parameters[25]. Details of the modeling may be found in the Methods section. Figure 4 shows the basic circuit, highlighting the main branches through which current is divided after the constriction switches: the shunt resistor (Fig. 4b), the constriction itself (Fig. 4c), and the loop inductor (Fig. 4d). In a device with an unshunted constriction (represented here as $R_s$ = 1 MΩ for consistency), nearly all of the bias current is diverted to the loop inductor after a switching event, causing the maximum amount of persistent current ≈ $I_c$ to be stored in the loop once the hotspot collapses and the constriction regains the superconducting state. In contrast, shunting the constriction with $R_s$ = 5 Ω allows the majority of the bias current to be diverted instead to the resistor, minimizing the amount of flux trapped through the loop inductor. Figure 4 b(i)-d(i) show the result over a longer timescale, where continuous switching of the shunted constriction brought on by a steadily increasing bias current ramp adds flux to the loop in increments of 5 $\Phi_0$, thus confirming the experimental observations displayed in Fig. 3 for Device 1.

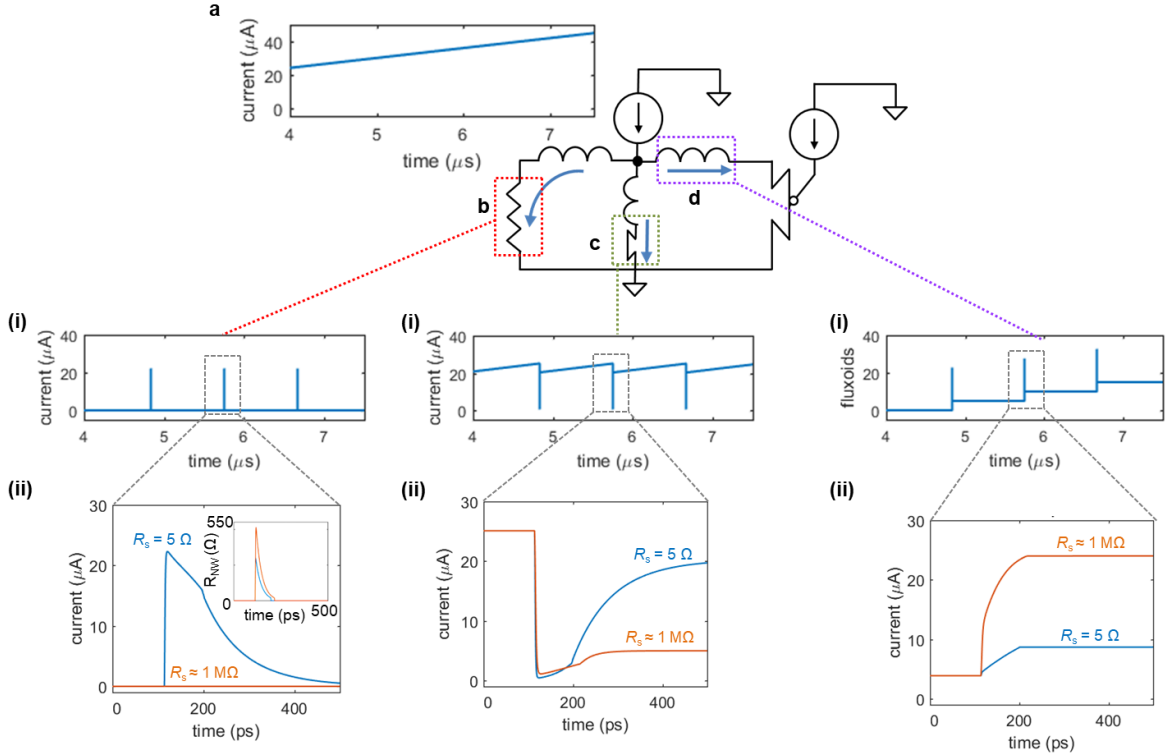

**Figure 4: Time domain simulations of the circuit, highlighting the three branches through which the bias current is diverted.** (a) Bias current ramp delivered to the device. (b) Current through the shunt resistor. *(i) Simulation of over a long time domain. (ii) Simulation over a single switching event. Time on the x-axis has been shifted to start from t = 0. In the 1 MΩ case, essentially no current is diverted to the resistor, and the maximum amount of current is stored in the loop. When $R_s$ = 5 Ω, nearly all of the current is diverted to the resistor, reducing the amount of trapped flux. Inset shows that the low shunt resistor also reduces the hotspot resistance and allows it to collapse more quickly.* (c) Current through the constriction. *(i) Simulation of over a long time domain. (ii) Simulation over a single switching event.* (d) Current through the loop inductor. *(i) Simulation of over a long time domain. The amount of flux trapped in the loop increases by 5 $\Phi_0$ every time the constriction switches. (ii) Simulation over a single switching event.* For all of these simulations, $L_{shunt}$ = 50 pH and $L_{loop}$ = 1.87 nH.

Figure 5 displays the amount of flux per switching event resulting from simulating devices of varying circuit parameters. Figure 5a shows that the amount of flux increases proportionally with increasing shunt resistance and shunt inductance, which agrees with the experimentally observed shift caused by increasing $R_s$. Figure 5b suggests a slightly more complex relationship between $R_s$ and $L_{loop}$, with plateaus occurring due to limitations on the maximum loop current with respect to the critical current of the constriction—e.g. if $I_c$ = 20 μA, a loop inductance leading to a ratio of 2 μA of circulating current per fluxon cannot have more than 10 fluxons per switching event. While both of these results rely on the electrothermal dynamics included in the simulation, their general shapes stem from current division between the shunt impedance and the loop impedance after the nanowire switches. Details on this relationship may be found in the Supplemental Material. In Figure 5c, the simulated trends for varying $R_s$ with $L_{loop}$ = 1.85 nH or $L_{loop}$ = 0.65 nH are compared to the experimental results for the three devices listed in Table 1. Data points representing the three measured devices show that the electrothermal simulations are in good agreement with the experimental measurements.

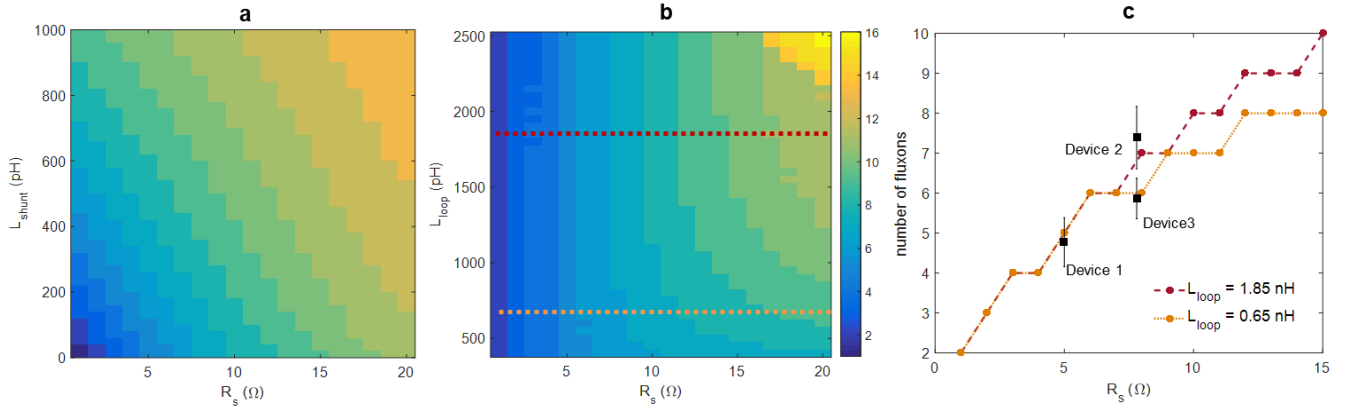

**Figure 5: Simulated effect of circuit parameters on amount of trapped flux.** (a) Number of fluxons per switching event as a function of varying $R_s$ and $L_{shunt}$. $R_s$ was swept in increments of 1 Ω, and $L_{shunt}$ was swept in increments of 50 pH. $R_s$ was $L_{loop}$ was held constant at 1.87 nH. (b) Number of fluxons per switching event as a function of varying $R_s$ and $L_{loop}$. $R_s$ was swept in increments of 1 Ω, and $L_{loop}$ was swept in increments of 50 pH. $L_{shunt}$ was held constant at 50 pH. (c) Comparison of the simulated number of fluxons per switching event with the three experimentally measured devices. The orange curve represents the trend for $L_{loop}$ = 0.65 nH, and the red curve represents the trend for $L_{loop}$ = 1.85 nH. Experimental mean values are represented as black squares. The error bars are $\pm$ 0.5 $\sigma$.

## 5. DISCUSSION ON APPLICATIONS

The results of our experiments and correspondence with simulations demonstrate that the device output may be tuned through simple circuit parameters and tailored to meet specific design requirements. As a result, the device is a promising platform for the development of a multilevel memory, with the number of states dictated by the critical current of the constriction and the amount of flux per event. While we demonstrated a maximum of 7 states in the case of Fig. 3, the simulations of Fig. 5 show that the number could be increased by changes that reduce the number of fluxons per event, such as further reducing $R_s$. This multilevel operation is significantly different from previously reported superconducting nanowire-based memories, which have thus far been predominantly binary devices[16 26 27]. It has recently been argued that multilevel memory may compensate for the large power consumption of peripheral circuits in superconducting memory arrays by providing a higher information capacity per cell given the same peripheral circuitry; additionally, multilevel memories may allow for increased memory density as the limits of physically shrinking a memory unit are reached[28]. Thus, future investigation of the device presented here as a multilevel memory may advance the scaling of superconducting memory arrays. Other potential applications of the device include use in multilevel logic or integration with other superconducting elements such as photon detectors. While the proof-of-concept device reported here had a rather large size (3 μm x 25 μm), the device could be scaled down by introducing a high kinetic inductance wiring layer for the loop, given that the geometric inductance is inconsequential. Further scaling improvements could be made by fabricating the loop as a stacked structure, as was suggested with previous nanowire-based memories[16].

## 6. CONCLUSIONS

In summary, we have developed a superconducting nanowire-based device capable of generating a response that is proportional to the strength of its input. By introducing local resistive shunting through on-chip fabrication, we are able to control the dynamics of a shunted constriction with a high-inductance superconducting loop and display behavior vastly different from its unshunted counterpart. When subjected to a DC bias current ramp, the device produces 7 distinguishable states as a result of controlled flux trapping, illustrating that it is able to be used as a multilevel memory. Through electrothermal circuit simulations and experimental measurements of devices with different circuit parameters, we showed how the amount of flux added or subtracted per event—and thus the number of distinguishable states—can be adjusted through device design. We envision that this device can be used as a tunable element for proportional and multilevel operations.

## 7. METHODS

**Device fabrication**

The shunt resistors and alignment marks were first patterned using electron-beam lithography (Elionix F125). A bilayer resist process was employed by first spinning the polymethyl methacrylate (PMMA) copolymer EL6 (6% in ethyl lactate) at 5 krpm for 60 s, then spinning the positive-tone resist gL2000 (Gluon Lab LLC) at 6 krpm for 60 s. The resist was developed in o-xylene and MIBK:IPA in a 1:3 ratio. A 10 nm Ti + 25 nm Au was evaporated, and lift-off was achieved in NMP heated to 60°C for 1 hour. A ~20-nm-thick NbN film was then deposited in an AJA sputtering system following the procedure described in [29]. The resulting sheet resistance was 150 Ω/sq, and the critical temperature was 8.5 K. The nanowire structures were then patterned with electron-beam lithography using gL2000, followed by cold development in o-xylene at 5 °C and reactive ion etching in $CF_4$ (Plasmatherm, RF power of 50 W, chamber pressure of 10 mTorr). The structure was imaged using a scanning electron microcrope (Zeiss) to check for proper alignment. The full process flow may be found in the Supplemental Material.

**Experimental setup**

All measurements were performed with the devices submerged in liquid helium at 4.2 K. The devices were adhered to a printed circuit board (PCB), and electrical connections between the devices and gold PCB pads were made using aluminum wire bonds. The PCB ports were connected to room-temperature electronics outside of the liquid helium dewar through CMP cables. Current-voltage characteristics were measured by applying a sinusoidal bias current from an arbitrary waveform generator (Agilent AWG33622A) at a sweep frequency of 10-20 Hz with a 10 kΩ series resistor. The DC output voltage was read by a 2 GHz, real-time oscilloscope (LeCroy 620Zi) after amplification through a low-noise preamplifier (Stanford Research Systems SRS560). The switching current of the yTron was measured by applying a voltage pulse through a 30 dB attenuator to the device, and measuring the skew between the oscilloscope trigger rising edge and the time at which a voltage output from the yTron was recorded, signifying a switching event. The bias pulse had a frequency of 500 Hz, width of 650 μs, rising edge of 400 μs, and height of 560 mV. The yTron output was sent through a pulse splitter and a low-noise amplifier (RF Bay LNA-2000, bandwidth: 10 kHz-2000 MHz, gain: 26 dB) before being read by an oscilloscope. The skew was then converted to units of switching current based on the slope of the bias waveform. To apply a DC bias to the constriction, a DC battery source (Stanford Research Systems

SIM928) was connected to the constriction through a 100 kΩ series resistor and a DC-1.9 MHz coaxial low-pass filter (MiniCircuits). To apply a pulse to the constriction, a pulsed voltage waveform of width ranging from 5 ns – 100 µs and height ranging from 50 mV- 550 mV was sent to the constriction input through a 30 dB attenuator. Details of the complete measurement setup may be found in the Supplemental Material.

**Statistics for flux per switching event measurements**

To measure the amount of trapped flux per step in bias current for each device, the bias current was ramped to ± 50 µA in increments of about 10% of the amount of current per fluxoid, or $0.1\Phi_0 L$, where $L$ is the total inductance of the loop. For $L_{loop}$ = 1.87 nH, $\Delta I_{bias}$= 0.1 µA, and for $L_{loop}$ = 0.66 nH, $\Delta I_{bias}$= 0.25 µA. At each bias point, the bias current was applied for 100 ms, and subsequently turned off. The device was allowed to rest for 100 ms before 10 readings of the yTron switching current were measured. Eight complete ramping cycles were recorded for each device. The bias current at the first point of each step was recorded, and the difference between sequential steps was calculated.

**Simulation details**

The circuit simulator used in this work uses a superconducting nanowire model implemented in Matlab that includes thermoelectric dynamics of hotspot formation and decay[24]. Physical parameters for the material stack are derived from prior literature[25] and are adjusted to match the experimental results. Flux quantization in the superconducting loop is enforced following each transient, which is found to be sufficient in explaining behavior in the absence of coherent transport events.

**Data availability**

The data that support the plots within this paper and other findings of this study are available from the corresponding author upon reasonable request.

**Code availability**

The code that support the plots within this paper and the electrothermal simulations used to confirm the findings of this study are available from the corresponding author upon reasonable request.

**ACKNOWLEDGEMENTS**

The authors thank Di Zhu, Andrew Dane, Dr. Reza Baghdadi, and all members of the Quantum Nanostructures and Nanofabrication Group for scientific discussions. They also thank Prof. Qing-Yuan Zhao for experimental advice and help with interpretation of results. The authors are grateful for James Daley and Mark Mondol of the MIT Nanostructures Laboratory for their technical support. This research is primarily supported by the Intel Corporation. Adam McCaughan was supported by a fellowship from the National Research Council. E.T. was supported by the National Science Foundation Graduate Research Fellowship Program (NSF GRFP) under Grant No. 1122374. Additional support for simulations and fabrication came from the Cryogenic Computing Complexity (C3) program and the DARPA Detect program through the Army Research Office under cooperative agreement number W911NF-16-2-0192. This research is based in part on work supported by the Office of the Director of National Intelligence (ODNI), Intelligence Advanced Research Projects Activity (IARPA), via Contract No. W911NF-14-C0089. The views and conclusions contained herein are those of the authors and should not be interpreted as necessarily representing the official policies or endorsements, either expressed or implied, of the ODNI, IARPA, or the U.S. Government. The U.S. Government is authorized to reproduce and distribute reprints for governmental purposes notwithstanding any copyright annotation thereon. This is an official contribution of the National Institute of Standards and Technology; not subject to copyright in the United States.

**AUTHOR CONTRIBUTIONS**

E.T. designed the device. E.T. and M.C. fabricated the devices. M.C. deposited the superconducting film. M.O. designed and performed the simulations. E.T., A.M., and K.K.B. conceived the experiments. E.T. performed the measurements with assistance from B.B. and M.C., and E.T. analyzed the data. E.T. wrote the manuscript with input from all of the authors. K.K.B. supervised the project. All of the authors contributed to discussions and improved understanding of the results.

# SUPPLEMENTAL MATERIAL

## Table of Contents

# I. Device fabrication

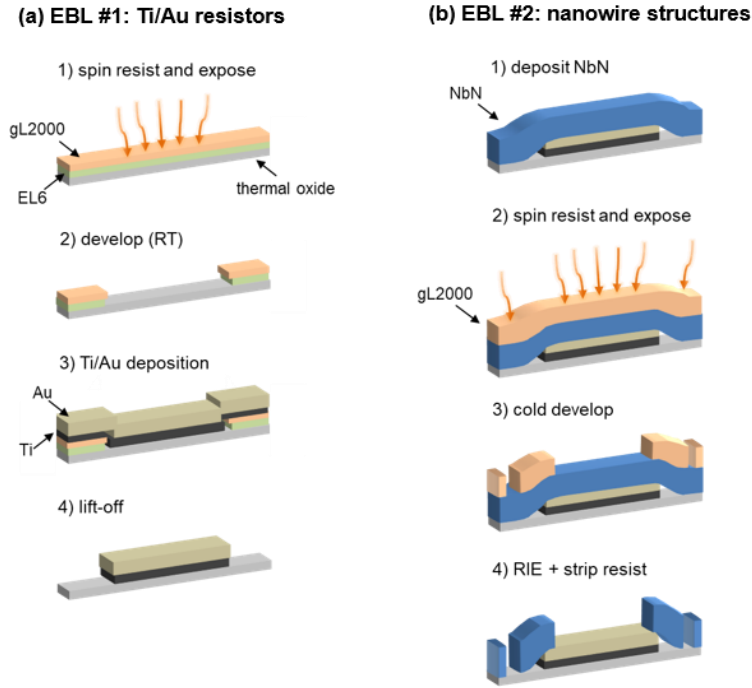

**Figure S1: Fabrication process.** *(a) Fabrication of the shunt resistors and alignment marks:* 1) EL6 was spin coated onto the sample at 5 krpm for 1 min and baked at 180 °C for 3 min. gL2000 was then spin coated onto the sample at 6 krpm for 1 min and baked at 180 °C for 3 min. A 125 keV electron-beam lithography system (Elionix F123) was used to expose the resist at a beam current of 5 nA, dose = 600 μC/cm$^2$. 2) The resist was first developed in o-xylene for 30 s and then developed in a MIBK:IPA (1:2 solution) for 1 min 40 s. 3) The metal layers (10 nm Ti/25 nm Au) were deposited using electron-beam evaporation. 4) Lift-off of the metal resistors was achieved using heated NMP at 60°C for 1 hr. *(b) Fabrication of the nanowire devices and windows over the resistors:* 1) NbN was deposited using an AJA sputtering system. 2) gL2000 was spun at 5 krpm for 1 min and baked at 180 °C for 2 min. The resist was exposed in the EBL system. The small nanowire features were written with a 500 pA beam current, and the larger pads were written with a 10 nA beam current. Both used a dose of = 600 μC/cm$^2$. 3) The resist was developed in o-xylene at 5 °C for 30 s, and IPA for 30 s. 4) The pattern was transferred to the underlying NbN using a reactive ion etch in CF$_4$ (power = 50 W, chamber pressure = 10 mTorr) for 4.5 min. The resist was removed by placing the chip in heated NMP at 60 °C for 1 hr, and leaving it in room temperature NMP overnight.

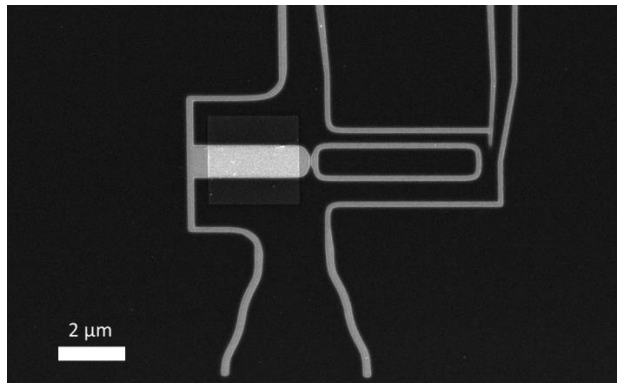

**Figure S2: Scanning electron micrograph of a device with smaller loop inductance.** Reducing the loop size led to a loop inductance of ~0.66 nH.

## II. Measurement details

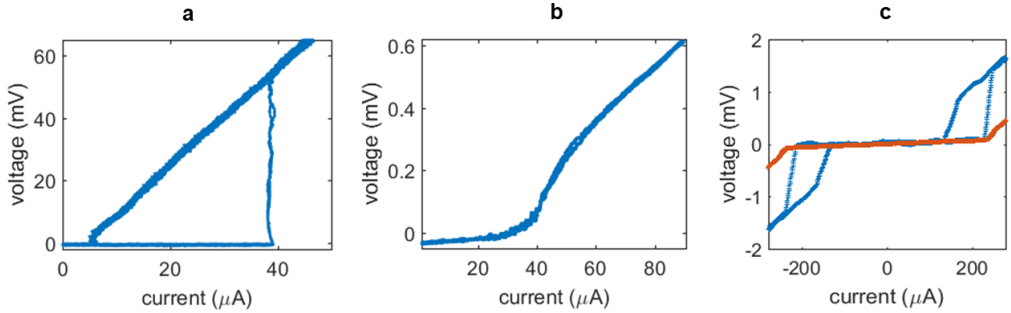

**Figure S3**: **Current-voltage characteristics of shunted and unshunted devices.** (a) Current-voltage characteristics of an unshunted 60-nm-wide nanowire. (b) Current-voltage characteristics of an identical nanowire, shunted with 5 Ω. (c) Comparison of the current-voltage characteristics of the complete device from the constriction input. The blue curve is the device shunted with ~ 7.8 Ω, and the red curve is the device shunted with ~ 5 Ω. The comparison shows that hysteresis for switching both sides of the superconducting loop is fully suppressed in the device with the 5 Ω shunt.

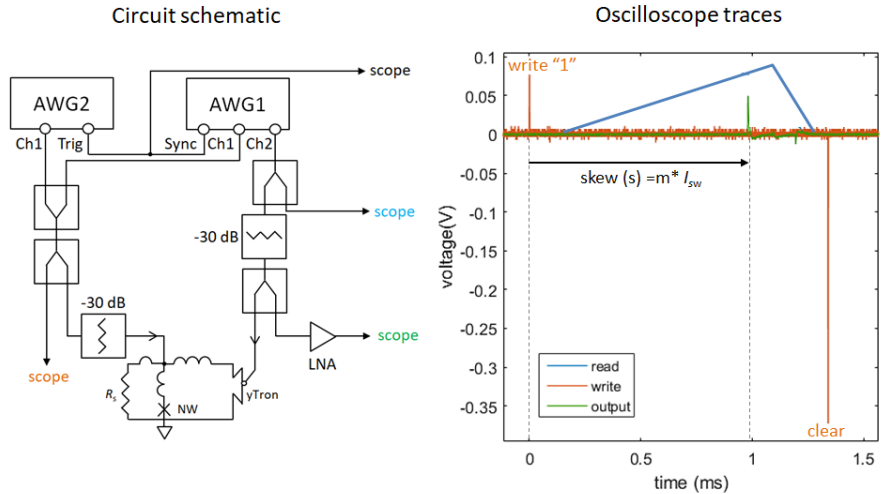

**Figure S4: Experimental setup for measuring the response of the device to a pulsed input.** Circuit schematic of the setup, and corresponding oscilloscope traces. Channel 2 of a dual channel arbitrary waveform generator, AWG1, (Agilent AWG33622A) was used to generate a voltage bias ramp for the right arm of the yTron. The pulse was sent to a power splitter, with one output feeding into the oscilloscope (blue trace) and the other output feeding into a second power splitter through a 30 dB attenuator. One output of the second power splitter was connected to the yTron port, and the second output was connected to a low-noise amplifier (RF Bay, LNA-2000) in order to read response of the yTron (green trace). A voltage pulse from the LNA occurred when the applied bias current exceeded the switching current of the yTron, causing it to switch. Channel 1 of AWG1 used to generate a writing pulse to the constriction, which combined with a negative "resetting" pulse from a second AWG (Agilent AWG 33250A) through a power splitter. The output of the power splitter fed into another power splitter; one of the outputs was sent to the oscilloscope (orange trace), while the other output was connected to the constriction through a 30 dB attenuator. The sync output of AWG1 is used to trigger AWG2, and is also sent to the oscilloscope to measure the skew from the rising edge of the trigger to the time at which the yTron output switches. The raw traces are later used to convert skew measurements into switching current.

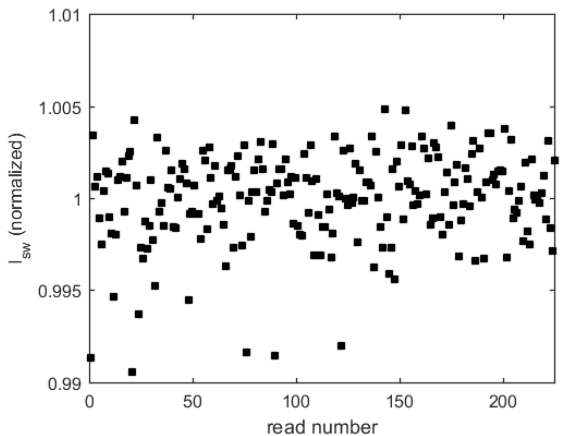

**Figure S5: Stability of the loop readout.** Plot shows 225 reads using a read pulse at 500 Hz with a 400 µs ramp time. The mean value is stable across all of the reads, indicating that the reading operation is nondestructive. The switching currents have been normalized to the mean of all the 225 values.

## III. Experimental details

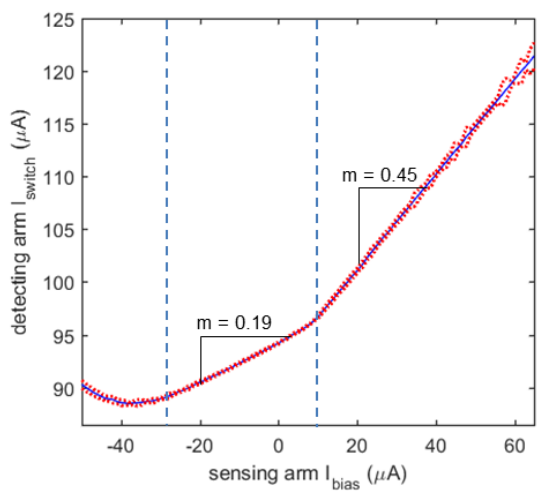

**Figure S6: Example of a yTron sensitivity curve showing different regions of sensitivity.** The plot shows the switching current of the detecting arm versus the amount of bias current sent through the opposite arm. Both arms were 300 nm wide. The dark blue trace shows the mean value of 100 sweeps; the red dashed lines are +/- one standard deviation from the mean. The slope changes from m = 0.19 to m = 0.45, indicating that the sensitivity curve is nonlinear.

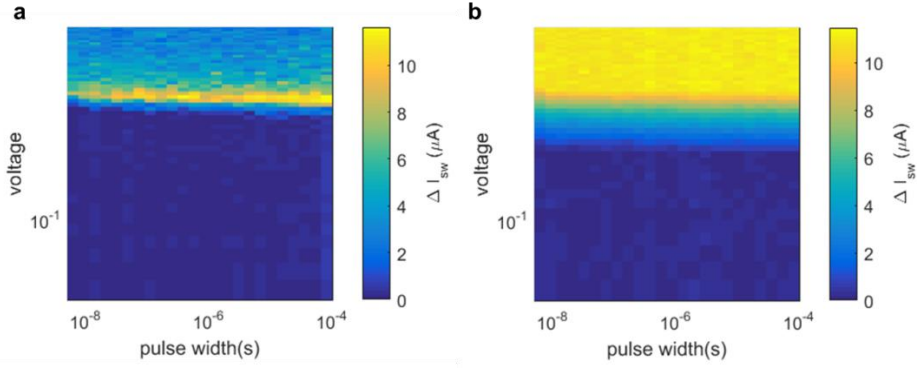

**Figure S7: Response of the devices to the input signal's voltage and pulse width.** (a) Logarithmic colormap showing the switching current of the right yTron arm as a function of the input voltage and pulse width for a superconducting loop with an unshunted constriction. The switching currents are plotted in terms of $\Delta I_{sw} = I_{sw} - I_{sw}(v = 0)$, where $v$ is the voltage height of the input pulse. (c) Colormap of the same measurement repeated on a superconducting loop with a shunted constriction, $R_s = 5\ \Omega$ (Device 1 in Table 1 of the main manuscript).

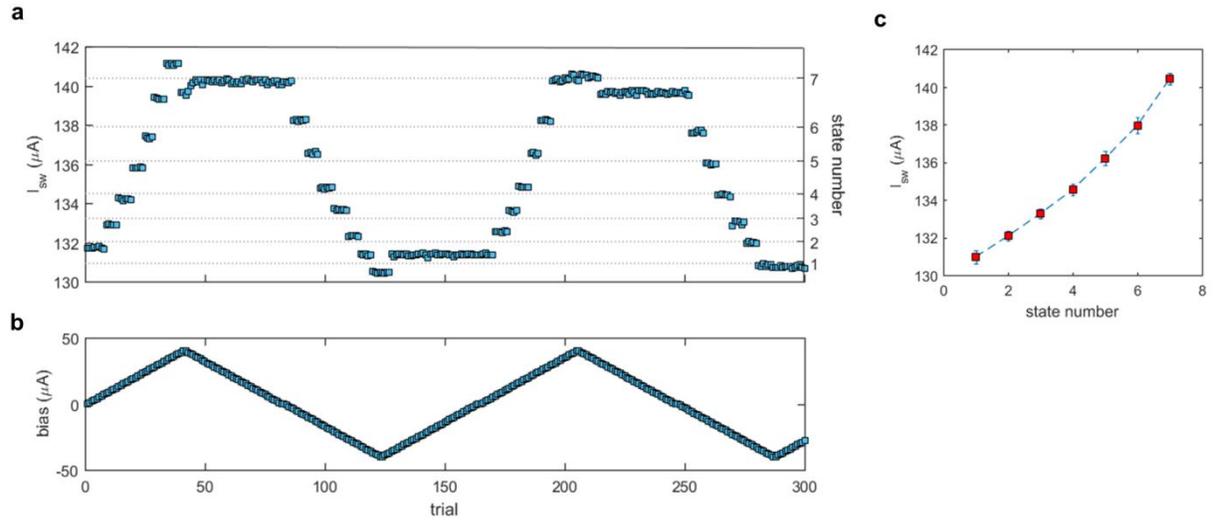

**Figure S8: yTron switching current measured over a DC bias ramp for Device 1 of Table 1.** . (a) Switching current of the right arm of the yTron in response to a DC bias ramp. Each point is the mean of 10 measurements of the yTron switching current. Each bias current was applied to the constriction for 100 ms before being turned off during the reading operation. (b) Bias current applied to the shunted constriction at each point of the plot in (a). (c) The seven consistent states observed in the yTron output of (a). The red marker is the average of eight state measurements, corresponding to four ramp cycles. The errorbars are the standard deviation of the eight measurements, where the standard deviation $\sigma = \sqrt{\sum_{i=1}^{n} \frac{|x_i - \mu|^2}{n-1}}$ .

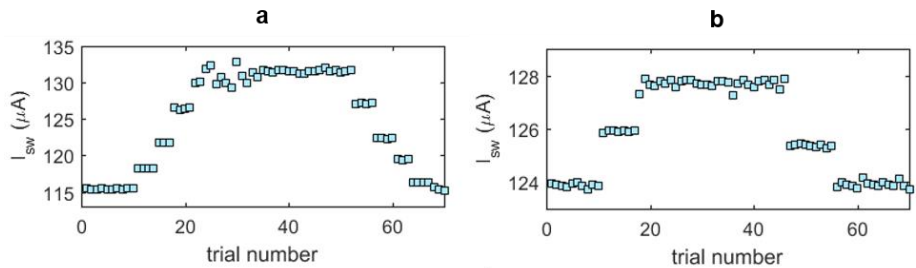

**Figure S9: yTron switching current measured over a DC bias ramp for Device 2 and Device 3 of Table 1**. (a) Results for Device 2 with a single bias ramp of ± 40 µA, showing 5 consistent states. (b) Results for Device 3, showing 3 consistent states.

## IV. Discussion of noise

In order to understand why we did not observe an average number of fluxons per step that was an exact integer (e.g. 4.77 $\Phi_0$ instead of 5 $\Phi_0$), we estimated a principle noise source of our experiment: the noise in measuring the yTron switching current using the oscilloscope. As shown in Figure S4, we measured the yTron switching current by recording the skew between the trigger and the point at which the yTron switched over the bias ramp, indicated by a voltage pulse. The skew was measured using the built-in function on the LeCroy oscilloscope. We then translated the skew into units of bias current using the ramp rate and taking into account the attenuators and 50 Ω impedances of the components. The baseline signal at 0 V had a standard deviation of 2.4 mV, corresponding to a skew noise of 4.82 µs with respect to the ramp rate of the yTron bias. Converting into units of bias current, this produces approximately 1.54 µA of noise in the yTron signal. This is the same order of magnitude as the standard deviation in average number of fluxons per step shown in Table 1 of the main manuscript, indicating that it is a significant contributor to the variation we observed experimentally.

Another contributor to the variation in the average number of fluxons per step is the behavior of the storage loop when it contained nearly its maximum amount of circulating current (~ $|I_{c,NW}|$). As shown in Figures S8 and S9, when the loop stored nearly its maximum amount at the plateaus of the bias ramp, it could lose a small amount of flux (~1-2 $\Phi_0$) in order to stabilize at high bias currents. This caused some variation in the position of the states of the output, as well as some variation in the amount of flux per step in the response. As a result, the average amount of flux per step was not an exact integer. It should be noted that this effect could be minimized by reducing the amplitude of the bias ramp, preventing the loop current from reaching its maximum amount. However, we have shown the result of the higher bias ramp in this work for clarity of the underlying mechanism in the devices.

## V. Flux dependence derived from basic current division

An intuitive sense of the trends shown in Figure 5 of the main text can be gained from simply calculating the flux predicted by current division between the shunt resistor and the loop inductor after the nanowire switches. The total impedance for the shunt resistor is the sum of the shunt resistance and the impedance of the shunt inductance, $Z_{shunt} = R_{shunt} + 2\pi f L_{shunt}$, while the impedance for the loop inductor is $Z_{loop} = 2\pi f L_{loop}$. The current charging the loop inductor, which dictates the amount of flux trapped after the loop heals, is $I_{loop} \cong I_c * \frac{Z_{shunt}}{Z_{shunt}+Z_{loop}}$, with the total number of trapped fluxons then being $I_{loop} L_{loop} / \Phi_0$.

The expression for loop current above shows that the problem reduces to the general form $z = \frac{x}{x+y}$. The shape of the trend for varying the loop inductance and the shunt resistance can be predicted by varying both variables, while the trend for varying the shunt inductance and shunt resistance only involves varying *x*. Figure S10 shows the analytical solution for varying both *x* and *y*.

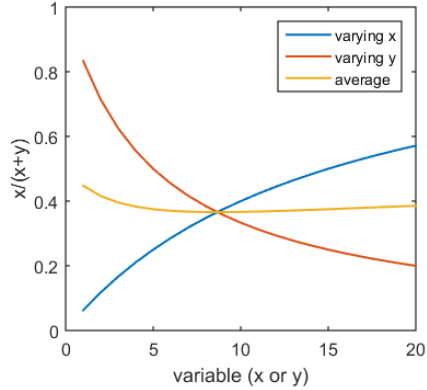

**Figure S10: Varying both variables in the general expression for loop current.** The variable *x* represents the shunt impedance, while the variable *y* represents the impedance of the superconducting loop. For varying *x*, *y* was set constant to 15. For varying *y*, *x* was set constant to 5.

Figure S11 shows the result of varying $R_{shunt}$ and $L_{shunt}$ (Fig. S11a) and varying $R_{shunt}$ and $L_{loop}$ (Fig. S11b) in the calculation of trapped flux above. In this case, $I_c = 25$ μA and *f* was arbitrarily set to 1.8 GHz. The shapes of the trends correspond to those in the simulations of Fig. 5 in the main text; however, it is clear that the electrothermal factors included in the more complex simulations of Fig. 5 are needed to tailor these trends to reflect the true timing dynamics of the nanowire and the thermal effects of the shunt resistor on the relaxation time and hotspot growth. Nonetheless, the general shape of the trends is clearly a result of the current division between the two impedances.

The code used to generate these figures is available upon request.

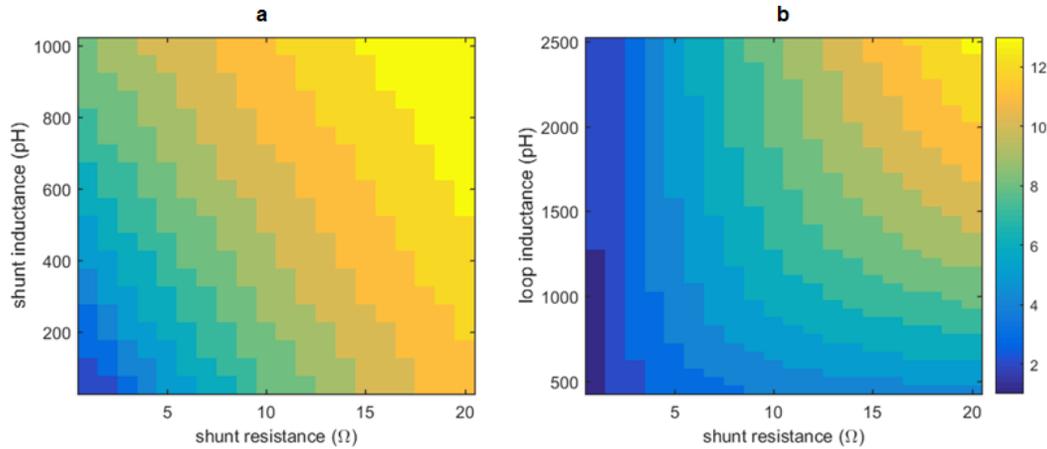

**Figure S11: Trapped flux dependencies calculated from simple bias current division.** (a) Varying the shunt resistance and shunt inductance. The loop inductance was set constant at 1.87 nH. (b) Varying the shunt resistance and the loop inductance. The shunt inductance was set constant at 50 pH.